\documentclass[intlimits,twoside,a4paper]{article}

\usepackage{amsmath,amssymb}
\usepackage{graphicx}
\usepackage{wrapfig}

\usepackage[T2A]{fontenc}
\usepackage[cp1251]{inputenc}

\usepackage[eqsecnum]{cmpj2}

\issue{2013}{16}{2}{23703}

\doinumber{10.5488/CMP.16.23703}

\newcommand{\non}{\nonumber \\}

\newcommand{\lp}{\left (}
\newcommand{\rp}{\right )}
\newcommand{\lb}{\left [}
\newcommand{\rb}{\right ]}
\newcommand{\lbr}{\left \{}
\newcommand{\rbr}{\right \}}
\newcommand{\be}{\begin{equation}}
\newcommand{\ee}{\end{equation}}
\newcommand{\bea}{\begin{eqnarray}}
\newcommand{\eea}{\end{eqnarray}}

\newcommand{\vR}{\vec{R}}

\newcommand{\vk}{\vec{k}}
\newcommand{\vl}{\vec{l}}
\newcommand{\cM}{{\cal{M}}}

\title[Ferroelectric order parameter]%
{Ferroelectric order parameter in the two-particle cluster system near phase transition point: collective variables method}
\author[M.A. Korynevskii, V.B. Solovyan]{M.A. Korynevskii\refaddr{label1,label2,label3},
        V.B. Solovyan\refaddr{label1}}
\addresses{
\addr{label1} Institute for Condensed Matter Physics of the National Academy of Sciences of Ukraine, \\1 Svientsitskii St., 79011 Lviv, Ukraine
\addr{label2} Lviv Polytechnic National University, 12 Bandera St., 79013 Lviv, Ukraine
\addr{label3} Institute of Physics of the University of Szczecin, 15 Wielkopolska St., 70451 Szczecin, Poland
}

\authorcopyright{M.A. Korynevskii, V.B. Solovyan, 2013}
\date{Received February 28, 2013, in final form April 2, 2013}

\begin{document}

\maketitle

\begin{abstract}
A new approach in the investigation of the order parameter behaviour near ferroelectric phase transition point is suggested.
The short range and dipole interactions between particles are taken into account. The logarithmic corrections
and effective critical exponents are calculated and discussed.
\keywords phase transition, order parameter, critical indices, collective variables method
\pacs 76.30.K, 77.80Bh
\end{abstract}

\section{Introduction}

The present paper has been prepared on the occasion of M.P.~Kozlovskii's 60-th anniversary, a prominent physicist and specialist in the
phase transition theory. As a collaborator of
I.R.~Yukhnovskii and an adept of this collective variables method (CVM) M.P. Kozlovskii has made much in applying this method to the  investigation of the Ising model,
especially to the study of the role of external field in the critical
behaviour of thermodynamic functions.

The advantage of the CVM, as compared with other methods, is a
possibility to obtain not only the universal characteristics near the phase transition point
(critical exponents and critical amplitudes ratio) but
also non-universal ones (i.e., expressions for different thermodynamic
functions) \cite{1,2}. In those investigations, the Ising model and
the $n$-component model with a spherical-symmetric potential of
interparticle interaction were presented. More realistic systems
(i.e., combination of short and long range or non-symmetric potentials)
were not studied. The problem is in the choice of the basic distribution for
CV and the form of layer-by-layer integration in the CV-space. These
two circumstances change the shape of recursion relations for
coefficients of basic distribution and are responsible for the
critical properties of a system.

In papers \cite{3,4_1,4_2,4_3,5,6}, a major solution for some of these problems was
suggested. There are two main points to be underlined. The first
one: the
transformation of a real Hamiltonian
into quasi-diagonal form (like Ising model).
And the second one: the frequency dependent CV
introduction and a choice of non-spherical form of CV-layers for integration.
The latter is necessary to take into account the peculiarity
of dipole-dipole interparticle interactions.

The aim of this paper is to demonstrate the capabilities of CVM in the
calculation of a partition function for a system of interacting clusters
having a ferroelectric type of ordering. The main goals are to calculate the equation
of state and to study the role of the external electric field in the formation of
ferroelectric order parameter.

\section{Hamiltonian. Partition function functional}

The problem regarding the phase transition in ferroelectric cluster systems starts from the well-known Slater paper \cite{7}
on the KH$_2$PO$_4$ (KDP) crystal. The most fundamental progress in this direction was obtained
by Blinc \cite{8}, who was the first to consider not only short range interparticle interactions of
particles in several groups--clusters, but also long range interactions between particles
from different clusters and the effect of a transverse field (such as the tunneling motion
of a particle in the two-minima potential). The self-consistent field approximation for particles from different clusters was exploited to calculate the free energy of such a system. Naturally,
the classical critical behaviour of the cluster system was observed in such approximation.

In order to investigate the real critical behaviour of a cluster system, modern methods of phase transition
description should be used. Most of them are based on the Ising or Heisenberg model study.
So, the initial cluster Hamiltonian should be reduced to an Ising-like form. This problem was solved in \cite{3} using
generalized Hubbard-Stasyuk operators \cite{3,9}.

We start with the Ising-like form of cluster Hamiltonian of
ferroelectric system with two quasispin particles in every cell of
crystalline lattice:
\be
H = \sum^{2^{2f_0}}_{i=1} \left\{ \sum_{q=1}^{N} \Lambda_i Y^i(\vec R_q) +
\sum_{q,q'=1}^{N} V_i (\vec R_q, \vec R_{q'}) Y^i (\vec R_q) Y^i (\vec R_{q'})\right\}.
\ee
Here $Y^i(\vR_q)$ is a generalized Hubbard--Stasyuk operator in a $q$ site; $\Lambda_i$ is the energy of every cluster of $f_0$
particles (including short range interparticle interaction and external electric field $E$);
$V_i(\vR_q\vR_{q'})$ is an eigenvalue of the dipole interaction matrix (see \cite{3,10}).

Taking the first term in (2.1) as the basic state of the investigated system, the general form for the partition function functional
in CVM can be obtained. With the accuracy up to the fourth order in the exponential form we have:
\bea
Z &=& Z_0 \int \left[ \rd\rho_\lambda(\vk,\nu)\right]^N \exp \lbr \sum_{\lambda} \sum_{k,\nu} \frac{\beta}{2} \Phi_\lambda(\vk)
\rho_\lambda(\vk,\nu) \rho_\lambda(-\vk,-\nu)\rbr \non
&&{}
\times \int \left[ \rd \omega_\lambda (\vk,\nu)\right]^N \exp \left\{ \ri 2 \pi \sum_{\lambda} \sum_{k,\nu}
\lb \rho_\lambda (\vk,\nu)-\cM_\lambda(\vk,\nu)\rb \omega_\lambda (\vk,\nu)  \right. \non
&&
{}
- \frac{(2\pi)^2}{2} \sum_{\lambda} \sum_{k,\nu} \cM_{\lambda\lambda} (\vk,\nu,-\vk,-\nu) \omega_\lambda(\vk,\nu)\omega_\lambda(-\vk,-\nu)  \non
&&
{}
+ \frac{(\ri 2\pi)^3}{3!} \sum_{\lambda_1,\lambda_2,\lambda_3} \sum_{{k_1,k_2,k_3}\atop{\nu_1,\nu_2,\nu_3}}
\cM_{\lambda_1\lambda_2\lambda_3} (\vk_1,\nu_1,\vk_2,\nu_2,\vk_3,\nu_3) \omega_{\lambda_1}(\vk_1,\nu_1) %
 \omega_{\lambda_2}(\vk_2,\nu_2) \omega_{\lambda_3}(\vk_3,\nu_3)  \non
&&
{}
+ \frac{(2\pi)^4}{4!} \sum_{\lambda_1,\lambda_2,\lambda_3,\lambda_4} \sum_{{k_1,k_2,k_3,k_4}\atop{\nu_1,\nu_2,\nu_3,\nu_4}}
\cM_{\lambda_1\lambda_2\lambda_3\lambda_4} (\vk_1,\nu_1,\vk_2,\nu_2,\vk_3,\nu_3,\vk_4,\nu_4) \omega_{\lambda_1}(\vk_1,\nu_1)\non
&&{}
\left.
 \vphantom{\sum_{\lambda_1}}
 \times \omega_{\lambda_2}(\vk_2,\nu_2) \omega_{\lambda_3}(\vk_3,\nu_3) \omega_{\lambda_4}(\vk_4,\nu_4) \rbr.
\eea
Here, $\rho_\lambda(\vk,\nu)$ is a CV, corresponding to $Y^i(\vR_q)$ operator in quasimomentum-frequency representation;
$\Phi_\lambda(\vk)$ is a Fourier transform of the intercluster dipole-dipole potential; $\beta = {1}/{kT}$, $k$ is the Boltzmann constant,
$T$ is the absolute temperature; $\cM_{\lambda}$, $\cM_{\lambda_1\lambda_2}$,  $\cM_{\lambda_1\lambda_2\lambda_3}$,
$\cM_{\lambda_1\lambda_2\lambda_3\lambda_4}$ are cluster cumulants of first, second, third and fourth order, $\omega_\lambda(\vk,\nu)$ are
variables conjugated to $\rho_\lambda(\vk,\nu)$; $Z_0$ is a partition function of non-interacting part of the Hamiltonian (2.1).
All rules of quasimomentum and frequency conservation in (2.2) are presented.

Since we investigate a two-particle cluster system ($f_0=2$) ($\lambda$ runs from 1 to 4), the explicit form of all coefficients in (2.2)
is as follows:
\be
Z_0 = Z_{01}^N = \lb \re^{\frac{\beta}{2}V} \lp 1 + 2 \cosh \beta E \rp + \re^{-\frac{3\beta}{2}V}\rb^N;
\ee
\bea
\cM_{1}(\vk,\nu) &=& \frac{\sqrt 2 \re^{\frac{\beta}{2}V} \sinh \beta E}{Z_{01}} \delta(\vk) \delta(\nu),\non
\cM_2(\vk,\nu) &=& 0,\non
\cM_{1 1}(\vk_1,\nu_1,\vk_2,\nu_2) &=& \left( \frac{\re^{-\frac{\beta}{2}V} \cosh \beta E}{Z_{01}} - \frac{2 \re^{\beta V}\sinh^2\beta E}{Z_{01}^2}\right)
\delta(\vk_1+\vk_2) \delta(\nu_1+\nu_2),\non
\cM_{2 2}(\vk_1,\nu_1,\vk_2,\nu_2) &=& \frac{4 V \re^{-\frac{\beta}{2}V} \sinh \beta V}{\beta(4V^2+\nu^2)Z_{01}}
\delta(\vk_1+\vk_2) \delta(\nu_1+\nu_2),\non
\cM_{1 1 1 }(\vk_1,\nu_1,\vk_2,\nu_2,\vk_3,\nu_3) &=& \frac{\re^{\frac{\beta}{2}V}\sinh \beta E}{\sqrt 2 Z_{01}}
\delta(\vk_1+\vk_2+\vk_3)\delta(\nu_1+\nu_2+\nu_3)  \non
&&\hspace{-2cm}{}
- \frac{3 \sqrt 2 \re^{\beta V} \sinh \beta E \cosh \beta E}{Z_{01}^2}
\delta(\vk_1+\vk_2) \delta(\vk_3)\delta(\nu_1+\nu_2)\delta(\nu_3)  \non
&&\hspace{-2cm}{}
+ \frac{4 \sqrt 2 \re^{\frac{3\beta}{2} V} \sinh^3 \beta E}{Z_{01}^3}
\delta(\vk_1) \delta(\vk_2) \delta(\vk_3)\delta(\nu_1)\delta(\nu_2)\delta(\nu_3), \non
\cM_{1 2 2}(\vk_1,\nu_1,\vk_2,\nu_2,\vk_3,\nu_3) &=& \frac{4 \sqrt 2 V \sinh \beta V \sinh \beta E}{\beta(4V^2+\nu^2)Z_{01}^2}
\delta(\vk_1+\vk_2) \delta(\vk_3)\delta(\nu_1+\nu_2)\delta(\nu_3),\non
\cM_{2 2 2}(\vk_1,\nu_1,\vk_2,\nu_2,\vk_3,\nu_3) &=& 0,\non
\cM_{1111}(\vk_1,\nu_1,\vk_2,\nu_2,\vk_3,\nu_3,\vk_4,\nu_4) &=& \frac{\re^{\frac{\beta}{2}V}\cosh \beta E}{2 Z_{01}}
\delta(\vk_1+\vk_2+\vk_3+\vk_4)\delta(\nu_1+\nu_2+\nu_3+\nu_4)  \non
&&\hspace{-2cm}{}
- \frac{4 \re^{\beta V}\sinh^2 \beta E}{Z_{01}^2}
\delta(\vk_1+\vk_2+\vk_3)\delta(\vk_4)\delta(\nu_1+\nu_2+\nu_3)\delta(\nu_4)  \non
&&\hspace{-2cm}{}
- \frac{3 \re^{\beta V}\cosh^2 \beta E}{Z_{01}^2}
\delta(\vk_1+\vk_2)\delta(\vk_3+\vk_4)\delta(\nu_1+\nu_2)\delta(\nu_3+\nu_4)  \non
&&\hspace{-2cm}{}
+ \frac{24 \re^{\frac{3\beta}{2} V} \sinh^2 \beta E \cosh \beta E}{Z_{01}^3}
\delta(\vk_1+\vk_2) \delta(\vk_3)\delta(\vk_4)\delta(\nu_1+\nu_2)\delta(\nu_3)\delta(\nu_4) \non
&&\hspace{-2cm}{}
- \frac{24 \re^{2\beta V} \sinh^4 \beta E}{Z_{01}^4}
\delta(\vk_1)\delta(\vk_2) \delta(\vk_3)\delta(\vk_4)\delta(\nu_1)\delta(\nu_2)\delta(\nu_3)\delta(\nu_4),\nonumber\\
\cM_{1122}(\vk_1,\nu_1,\vk_2,\nu_2,\vk_3,\nu_3,\vk_4,\nu_4) &=& - \frac{4 V \sinh \beta V \cosh \beta E}{\beta(4V^2+\nu^2)Z_{01}^2}
 \delta(\vk_1+\vk_2) \delta(\vk_3+\vk_4)\delta(\nu_1+\nu_2)\delta(\nu_3+\nu_4)  \non
&&\hspace{-2cm}{}
+ \frac{8 V \re^{\frac{\beta}{2}} V \sinh \beta V \sinh^2 \beta E}{\beta(4V^2+\nu^2)Z_{01}^3}
\delta(\vk_1+\vk_2) \delta(\vk_3)\delta(\vk_4)\delta(\nu_1+\nu_2)\delta(\nu_3)\delta(\nu_4),\non
\cM_{2222}(\vk_1,\nu_1,\vk_2,\nu_2,\vk_3,\nu_3,\vk_4,\nu_4) &=& \frac{3 \re^{-\frac{\beta}{2}V} \lb \cosh \beta V - \frac{4V \sinh \beta V}{\beta(4V^2+\nu^2)}
\rb}{2\beta^2(4V^2+\nu^2)Z_{01}}  \nonumber\\[1ex]
&&\hspace{-2cm}{}
\times \delta(\vk_1+\vk_2+\vk_3+\vk_4)\delta(\nu_1+\nu_2+\nu_3+\nu_4)  \nonumber\\[1ex]
&&\hspace{-2cm}{}
- \frac{48 V^2 \re^{-\beta V}\sinh^2 \beta V}{\beta^2(4V^2+\nu^2)Z_{01}^2} \delta(\vk_1+\vk_2) \delta(\vk_3+\vk_4)\delta(\nu_1+\nu_2)\delta(\nu_3+\nu_4),
\eea
where $V$ is a short range interparticle interaction constant.

Among two CV $\rho_1(\vk,\nu)$ and $\rho_{2}(\vk,\nu)$, the first one is related to the ferroelectric order parameter
\be
\langle \rho_{1}(0,0)\rangle = \frac{1}{2} \left[ \sigma_1^z(\vR_q) + \sigma_2^z(\vR_q) \right]
\ee
and the second one is related to the antiferroelectric order parameter
\be
\langle \rho_{2}(0,0)\rangle = \frac{1}{2} \left[ \sigma_1^z(\vR_q) - \sigma_2^z(\vR_q) \right],
\ee
where $\sigma_f^z(\vR_q)$ is a quasispin of $f$-th particle in the $q$ cluster.

In order to calculate the ferroelectric order parameter, the integration in (2.2) over all CV with the exception of $\rho_{1}(0,0)$
must be performed. As a result, the equation of state will be obtained.

\section{Integration in CV space}

Since the ferroelectric order parameter is determined by the mean value of $\rho_{1}(0,0)$, the effect of $\rho_{2}(\vk,\nu)$
is not crucial and integration in (2.2) over $\rho_{2}(\vk,\nu)$ may be performed in a lower approximation i.e. using
Gaussian distribution only. For this purpose, the higher orders of the products of $\rho_{2}(\vk,\nu)$ and $\omega_{2}(\vk,\nu)$ variables in (2.2)
must be decomposed into a series and every term should be represented as a Gaussian momentum.
Using the following mathematical trick  for $n$-order product of $\omega_\lambda(\vk,\nu)$ variables:
\bea
\lefteqn{
\exp \left\{ \ri 2 \pi \sum_{k,\nu} \rho_\lambda(\vk,\nu) \omega_\lambda(\vk,\nu) - \frac{(2\pi)^2}{2} \sum_{k,\nu} \cM_{\lambda\lambda} (\vk,\nu,-\vk,-\nu)
\omega_\lambda(\vk,\nu) \omega_\lambda(-\vk,-\nu) \right\}} \nonumber\\[1ex]
&&{}
\times  \omega_\lambda(\vk_1,\nu_1) \omega_\lambda(\vk_2,\nu_2) \ldots \omega_\lambda(\vk_n,\nu_n)  \non
&&{}
= \frac{1}{(2\pi \ri)^n} \frac{\partial^n}{\partial\rho_\lambda(\vk_1,\nu_1)\partial\rho_\lambda(\vk_2,\nu_2) \ldots \partial\rho_\lambda(\vk_n,\nu_n)}
\exp \left\{ \ri 2 \pi \sum_{k,\nu} \rho_\lambda(\vk,\nu)\omega_\lambda(\vk,\nu)  \right. \non
&&{}
\left. - \frac{(2\pi)^2}{2} \sum_{k,\nu} \cM_{\lambda\lambda}(\vk,\nu,-\vk,-\nu) \omega_\lambda(\vk,\nu)\omega_\lambda(-\vk,-\nu)\rbr,
\eea
reduces the integration over $\omega_2(\vk,\nu)$  to the calculation of a simple integral:
\be
I = \int \left[ \rd \omega_2(\vk,\nu)\right] \exp \left\{ \ri 2 \pi \sum_{k,\nu} \rho_2(\vk,\nu) \omega_2(\vk,\nu) -\frac{(2\pi)^2}{2} \sum_{k,\nu}
\cM_{22}(\vk,\nu,-\vk,-\nu) \omega_2(\vk,\nu)\omega_2(-\vk,-\nu)\rbr.
\ee
Now, the integration of (2.2) of real $\rho_2^{\mathrm{c}}(\vk,\nu)$ and imaginary  $\rho_2^{\mathrm{s}}(\vk,\nu)$ parts of
\be
\rho_2(\vk,\nu) = \rho_2^{\mathrm{c}}(\vk,\nu) + \ri \rho_2^{\mathrm{s}}(\vk,\nu)
\ee
can be easy performed.

As a result, the total partition function functional (2.2) of exclusively CV $\rho_{1}(\vk,\nu)$, which form the branch of the investigated
system active in the ferroelectric phase transition, is obtained:
\bea
Z &=& Z_0 Z_2 \int \left[ \rd \rho_{1}(\vk,\nu)\right]^N \exp \lbr \sum_{k,\nu} \frac{\beta}{2} \Phi_{1}(\vk)
\rho_{1}(\vk,\nu) \rho_{1}(-\vk,-\nu)  \rbr\non
&&{}
\times \int \left[ \rd \omega_{1}(\vk,\nu)\right]^N \exp \left\{ \ri 2 \pi \sum_{k,\nu} \lb \rho_{1}(\vk,\nu)  -
\cM_{1}(\vk,\nu)\delta(\vk)\delta(\nu)\rb  \omega_{1}(\vk,\nu)   \right. \non
&&{}
- \frac{(2\pi)^2}{2} \sum_{k,\nu} \left[ \cM_{11}(\vk,\nu,-\vk,-\nu) + \frac{1}{12}
\sum_{k',\nu'}\cM_{1122}(\vk',\nu',-\vk',-\nu',\vk,\nu,-\vk,-\nu) g_2(\vk',\nu')\right] \non
&&{}
\times \omega_{1}(\vk,\nu)\omega_{1}(-\vk,-\nu) + \frac{(2\pi)^4}{4!} \sum_{{k_1,\dots ,k_4}\atop{\nu_1,\dots ,\nu_4}}
\cM_{1111}(\vk_1,\nu_1,\dots ,\vk_4,\nu_4)%
 \left. \omega_{1}(\vk_1,\nu_1)\ldots \omega_{1}(\vk_4,\nu_4) %
 \vphantom{\sum_{k,\nu}}
 \rbr.
\eea
In (3.4)
\bea
Z_2 &=& \prod_{k,\nu} \lbr \left[ 1 - \beta \Phi_2(0) \cM_{22}(0,0) \right]^{-\frac{1}{2}} \left[ 1 - \beta \Phi_2(\vk) \cM_{22}(\vk,\nu) \right]^{-1}\rbr \non
&&{}
\times \exp \lbr \frac{1}{8} \sum_{k,\nu,k',\nu'} \cM_{2222}(\vk,\nu,-\vk,-\nu,\vk',\nu',-\vk',-\nu') g_2(\vk',\nu')\rbr
\eea
is a partition function of non-active in the ferroelectric phase transition subsystem with $\lambda=2$, and
\be
g_2(\vk,\nu) = \beta\Phi_2(k) \left[ 1 - \beta \Phi_2(\vk) \cM_{22}(\vk,\nu)\right]^{-1}
\ee
is a corresponding screening potential.

As far as the ferroelectric phase transition in its origin is a classical phenomenon, only CV with zero Matsubara frequencies change their distribution form in
the phase transition point from Gaussian to non-Gaussian. Thus, the summation over all $\nu\neq 0$
may be performed in the Gaussian approximation. As a result, using nodal $(l)$ variables:
\be
\rho_l = \frac{1}{\sqrt N} \sum_k \rho_{1}(\vk,0) \re^{\vk\vl},
\ee
and the fourth-order basic measure density for $\rho_l$, the working form for integration of (3.4) over $\rho_l$ can be obtained:
\be
Z = Z_0 Z_2 Z_{1}\,,
\ee
where
\be
Z_{1} = \sqrt 2^{N-1} Q \int \left[ \rd\rho_{1}(\vk,0)\right]^N \exp \lbr \frac{\beta}{2} \sum_k \Phi_{1}(\vk) \rho_{1}(\vk,0)\rho_{1}(-\vk,0)\rbr \prod_l \lbr - \frac{1}{2} a_2^{(1)} \tilde\rho_l^2 - \frac{1}{4!} a_4^{(1)}\tilde\rho_l^4\rbr;
\ee
\bea
\tilde\rho_l &=& \rho_l - \sqrt N \cM_{16}, \qquad Q = 2 \int_0^\infty f(\omega) \rd\omega,\non
a_2^{(1)} &=& (2\pi)^2 Q^{-1} \int_{-\infty}^\infty \omega^2 f(\omega) \rd\omega, \qquad
a_4^{(1)} = -(2\pi)^4 Q^{-1} \int_{-\infty}^\infty \omega^4 f(\omega) \rd\omega + 3(a_2^{(1)})^2,\non
f(\omega) &=& \exp \lbr - \frac{(2\pi)^2}{2}\tilde\cM_{11}\omega^2 - \frac{(2\pi)^4}{4!} \cM_{1111}\omega^4 \rbr.
\eea
The renormalized cumulant $\tilde\cM_{11}$ (obtained after summation over $\nu\neq 0$) takes the following form:
\bea
&&\!\!\!\!\!\!\!\!\!\!\!\!
\tilde\cM_{11}(\vk,0,-\vk,0) \!=\! \cM_{11}(\vk,0,-\vk,0) -
\frac{M}{24} \sum_{k'} \beta\Phi_2(\vk') \lbr \frac{1}{V-\beta\Phi_2(\vk')N} + \frac{\beta\sqrt V \coth \beta
\sqrt{V[V-\beta\Phi_2(\vk')]N}}{\sqrt{V-\beta\Phi_2(\vk')N}}\rbr,\non
&&\!\!\!\!\!\!\!\!\!\!\!\!
M = \frac{\sinh \beta V}{\beta Z_{01}^2} \lp \cosh \beta E - \frac{2\re^{\frac{\beta}{2}}\sinh^2 \beta E}{Z_{01}}\rp, \qquad
N = \frac{\re^{-\frac{\beta}{2}V}\sinh \beta V}{\beta Z_{01}}\,.
\eea
To integrate (3.9) over CV $\rho_{1}(\vk,0)$, we will use the Yukhnovskii's layer-by-layer method \cite{1,11}, which is based
on the substitution of $\Phi_{1}(\vk,0)$ by its mean value in a narrow range of $\vk$ in the Brillouin zone.
In fact, here the Fourrier transform of the dipole potential is presented:
\be
\Phi_{1}(\vk) = \varphi_0 - \lambda \cos^2 \Theta - A|\vk|^2,
\ee
where $\varphi, \lambda, A$ are some constants, $\tilde\Theta$ is a polar angle. The layers of integration are not spherically
symmetric as compared with the situation for a simple Ising model \cite{1,2}. The presence of the term $\lambda\cos^2\Theta$ in (3.12)
leads to the situation in which the coefficient $a_2^{(n)}$ in some subzone $B_n^{\mathrm{G}}$ in $n$-th layer $B_n$ becomes positive and the corresponding integral in
(3.9) is finite even in the Gaussian approximation. In $B_n^{\mathrm{q}} = B_n-B_n^{\mathrm{G}}$, the non-Gaussian distribution must be used
to fulfill the condition of convergency of the corresponding integrals. The sequence of zones $B_n$, $B_n^{\mathrm{G}}$, $B_n^{\mathrm{q}}$ is presented by the next relations:
\bea
&&
B_n:\qquad \lb 0<|\vk|\leqslant  \frac{B_1}{S^{n-1}}; \qquad \Theta_{n-1}< \Theta \leqslant  \pi-\Theta_{n-1}; \qquad o<\varphi\leqslant  2\pi \rb,\non
&&
B_n^{\mathrm{G}}:\qquad \lb 0<|\vk|\leqslant  \frac{B_1}{S^{n-1}}; \qquad \Theta_{n-1}< \Theta \leqslant  \Theta_{n}, \quad \pi - \Theta_n < \Theta \leqslant  \pi-\Theta_{n-1}; \qquad
o<\varphi\leqslant  2\pi \rb,\non
&&
B_n^{\mathrm{q}}:\qquad \lb \frac{B_1}{S^n}<|\vk|\leqslant  \frac{B_1}{S^{n-1}}; \qquad \Theta_{n}< \Theta \leqslant  \pi-\Theta_{n}; \qquad o<\varphi\leqslant  2\pi \rb,
\eea
where $B_1$ is the initial Brillouin zone, $\varphi$ is the azimuth angle, $S\geqslant  1$ is the parameter for dividing the Brillouin zone into layers \cite{1,10}:
\be
\Theta_n = \arctan \sqrt{\frac{\beta\lambda}{\beta\varphi_0-a_2^{(n)}}-1}.
\ee

The general scheme of the proposed two-stage integration for magnetic systems was presented in \cite{6}. Here, we present only the recursion relations
for coefficients of consequent basic distributions, which determine the critical character of fluctuation processes near $T_{\mathrm{c}}$
in a ferroelectric cluster system with dipole-dipole interactions.
\bea
r_{n+1} &=& S^2 (r_n+q) \lbr \bar N_n + \frac{3}{4z_n} \lp \bar N - \frac{1}{3} \rp \left[ 1 - S^{-n}\sqrt{\frac{3(-r_n)}{2\beta\lambda}}\right]\rbr - S^2q,\non
U_{n+1} &=& \sqrt{\frac{r_n}{r_{n-1}}} {\cal E}_n U_n\,.
\eea
Here,
\bea
&&
r_n = S^{2n}d_2^{(n)}, \qquad  d_2^{(n)} = a_2^{(n)} - \beta\Phi_{1}(\vk), \qquad  U_n = S^{4n} a_4^{(n)};\non
&&
q = \frac{3}{5} \frac{1-s^5}{1-s^3} \beta A B_1^2;\non
&&
\bar N_n = \frac{2\sqrt{\xi_n}K(\xi_n)}{3\sqrt{z_n}K(z_n)}+\frac{1}{3}\,, \qquad  {\cal{E}}_n = S^6 \frac{L(\xi_n)}{L(z_n)}\,,\non
&&
z_n = \frac{3[d_2^{(n)}]^2}{4a_4^{(n)}}\,, \qquad  \xi_n = \frac{3}{2} S^3 \frac{K^2(z_n)}{L(z_n)}\,, \non
&&
K(z_n) = \sqrt{z_n} \left[ \frac{K_{3/4}(z_n)}{K_{1/4}(z_n)}-1\right]; \qquad
L(z_n) = 6 K^2(z_n) + 4\sqrt{z_n} K(z_n) - 1;
\eea
$K_{1/4}(z_n)$, $K_{3/4}(z_n)$ are Bessel functions of $z_n$ argument.

At $T<T_{\mathrm{c}}$, the integration over $\rho_{1}(\vk,0)$ in (3.9) should be performed up to $n=\mu_\tau$, which is determined from the following relation:
\be
d_2^{(\mu_\tau)} (B_{\mu_\tau}) = 0
\ee
and it is equal to
\be
\mu_\tau = 1 + \ln \frac{C_2 R-q}{C_1}\Bigg/ \ln E_1\,.
\ee
Here, $E_1$ is the bigger of the two eigenvalues of the matrix of relations (3.15) linearized in the neighborhood of
their fixed point, $C_1\sim \tau\ln^{-1/3}|\tau|$, $\tau={(T_{\mathrm{c}}-T)}/{(T_{\mathrm{c}})}$, $C_2$, $R$ are constants independent of $n$.

The logarithmic corrections to $C_1$ arose when the Gaussian integration was performed in $B_n^{\mathrm{G}}$ subzones with the summation of infinite series of
diagrams (as it was proposed in paper \cite{12}). It must be noted that for a spherically-symmetric potential (i.e., simple Ising model), such corrections are neglected.
However, in nonspherically-symmetric case (for dipole-dipole interaction), they are important.

In order to correctly extract the CV $\rho_{1}(\vk,0)$ with zero values of quasimomentum (i.e., order parameter) in (3.9), the displacement of the center of its fluctuation in the ordered phase should be made:
\be
\rho_{1}(\vk,0) = \rho_{1}(\vk',0) + \sqrt{2N} \langle \sigma_1^z\rangle \delta(\vk).
\ee
As a result, the fourth-order form integrand in (3.9), for the last ($n=\mu_\tau$) stage of integration, transforms into
\be
E_{\mu_\tau}(\rho_{1}(0,0)) = \lb a_2^{(\mu_\tau)} + 2 \bar d_2(0) \rb \sqrt N \cM_{1} \rho_{1}(0,0)
+ \frac{1}{4} \lb \bar d_2(0) - a_2^{(\mu_\tau)} I\rb \rho_{1}^2(0,0) - \frac{1}{N} a_4^{(\mu_\tau)} \rho_{1}^4(0,0),
\ee
where
\be
\bar d_2(\vk) = 2 |d_2^{(\mu_\tau)}(0)| + qk^2, \qquad  I = \frac{1}{N_{\mu_\tau}} \sum_k \frac{1}{\bar d_2(\vk)} =
\frac{3(t-\arctan t)}{\bar d_2(0) t^3}\qquad t = \frac{\pi}{bS^{\mu_\tau}} \sqrt{\frac{q}{\bar d_2(0)}}\,,
\ee
$b$ is a lattice constant.

Finally, for functional integral in (3.9), the following expression has been obtained:
\be
Z_{\mu_\tau} = f(T) \int \exp \lbr N \lp A\rho + B \rho^2 - D\rho^4\rp \rbr \rd\rho.
\ee
Here, $\rho \equiv \rho_{1}(0,0)$, $f(T)$ is a continuous function of temperature in the phase transition point neighborhood and
\be
A = \lb a_2^{(\mu_\tau)} + \frac{\bar d_2(0)}{2}\rb \cM_{1}\,, \qquad B = \frac{1}{4} \lp \bar d_2(0) - a_4^{(\mu_\tau)} I\rp, \qquad
D = \frac{1}{4!} \frac{N}{N_{\mu_\tau}} a_4^{(\mu_\tau)}.
\ee
Thus, the procedure of integration in the partition function functional (3.4) over all CV is performed except the one-fold integral (3.22)
which determines the order parameter of the cluster ferroelectric system.

\section{Order parameter and dielectric susceptibility}

To obtain the explicit form for coefficients $A, B, D$ we had to use the solution of linearized recursion equations (3.15)
in the vicinity of the Gaussian type fixed point (3.15), (3.16).
\bea
&&
A = \left[ \beta \Phi_{1}(0) \ln^{-2/3}|\tau| + \Delta_1 \tau \ln^{-1/3} |\tau| \right] \frac{\sqrt 2 \sinh \beta h}{1 + 2\cosh \beta h + \re^{-2\beta V}}\,,\non
&&
B = \frac{\Delta_1}{4} \left( 1 - \frac{3\Delta_2}{\Delta_1} \frac{t - \arctan t}{t^3} \right) \tau \ln^{-1/3} |\tau|,\non
&&
D = \frac{c_2^2 R^2 \Delta_2}{6 \Delta_1} \ln^{-2/3}|\tau|.
\eea

In CVM, the coefficients $\Delta_1, \Delta_2$ do not depend on the form of interparticle potential $\Phi_{1}(\vk)$ and are the same as
for isotropic Ising model \cite{2}.

The minimization of the integrand in (3.22) with respect to the variable $\rho$ yields the equation of state:
\be
A + 2 B \rho - 4 D \rho^3 = 0.
\ee

The temperature dependence of the real roots of this equation at different values of the external field is presented in figure~1.
Here, we used the energy parameters $\Phi_{1}(0) = 265$~K, $V = 40$~K. One can see a different situation for $T<T_{\mathrm{c}}$ and $T>T_{\mathrm{c}}$. At $E=0$
there are three roots for $T<T_{\mathrm{c}}$ but only one root for $T>T_{\mathrm{c}}$. At $h=0$, only two symmetric non-zero roots for $T<T_{\mathrm{c}}$ remain.
\begin{figure}[htbp]
\centerline{\includegraphics[width=0.5\textwidth]{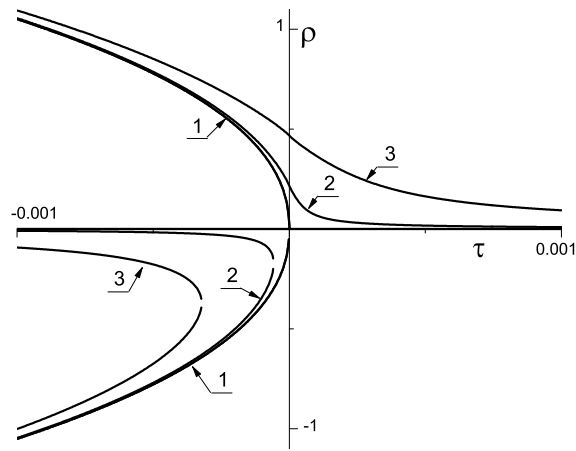}}
\caption{
Temperature dependencies of the real roots of equation of state (4.2) at different values of external field $E$: 1 -- $E = 0$~K,
2 -- $E = 0.1$~K, 3 -- $E=0.01$~K.
}
%%%\label{prsfig2}
\end{figure}

The field dependence of the real roots (4.2) for different temperatures is shown in figure~2. For $T>T_{\mathrm{c}}$, only a unique real root exists, but for
$T<T_{\mathrm{c}}$, there arise three real roots forming a non-synonymous situation. This problem is solved due to using the stability condition
$\left( {\partial \rho}/{\partial E}>0\right)$, so the hysteresis in the $\rho$ behaviour takes place in the region $-E_\tau\leqslant E \leq E_\tau$
(see figure~3). Naturally, the value of $E_\tau$ depends on temperature.
\begin{figure}[htbp]
\centerline{\includegraphics[width=0.5\textwidth]{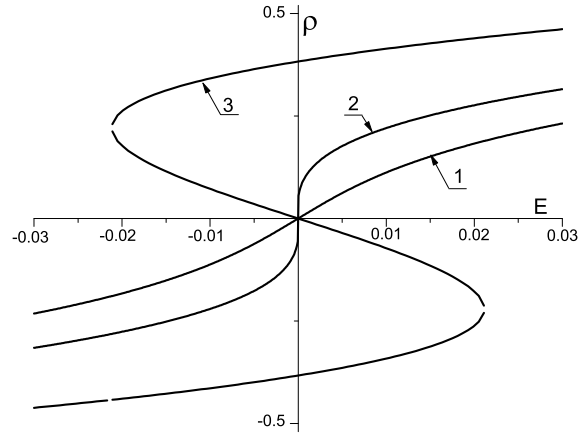}}
\caption{
%%ђЁб.~2.
Field dependencies of the real roots of the equation of state (4.2)
at different values of relative temperature $\tau$: 1 -- $\tau = 5\cdot 10^{-5}$,
2 -- $\tau = - 1 \cdot 10^{-6}$, 3 -- $\tau = - 1 \cdot 10^{-4}$.
}
%%\label{prsfig2}
\end{figure}
\begin{figure}[htbp]
\centerline{\includegraphics[width=0.5\textwidth]{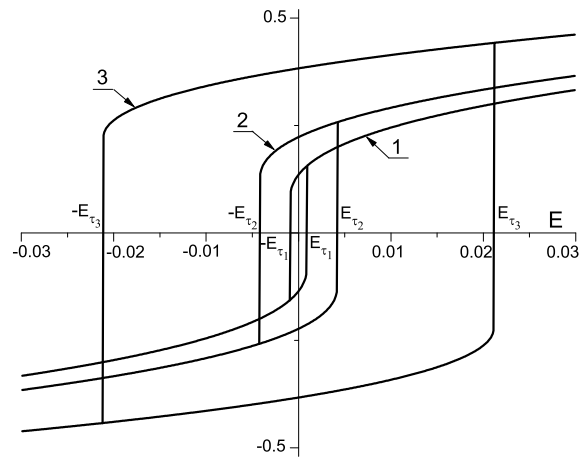}}
\caption{
%%ђЁб.~3.
Hysteresis loops of the order parameter for different values of the relative temperature $\tau$: 1 -- $\tau = -1 \cdot 10^{-5}$,
2 -- $\tau = -3 \cdot 10^{-5}$, 3 -- $\tau = -1 \cdot 10^{-4}$.
}
%%\label{prsfig2}
\end{figure}

A static dielectric susceptibility as a function
\be
\chi = \beta \left( \frac{\partial \rho}{\partial \beta E}\right)_{E = 0}
\ee
is shown in figures~4 and 5, where temperature depending ``susceptibilities'' are presented for different values of non-zero external
field (figure~4). The ``suppressive'' action and a shift of $\chi$ maximum are observed with an increase of $E$.
\begin{figure}[htbp]
\centerline{\includegraphics[width=0.5\textwidth]{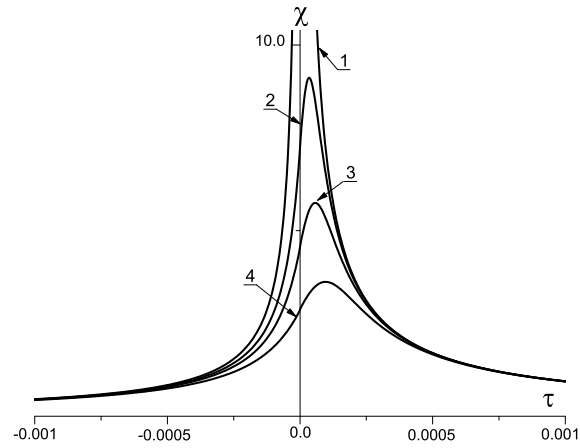}}
\caption{
%%ђЁб.~3.
Temperature dependencies of the static dielectric ``susceptibility'' for different values of the external field $E$:
1 -- $E = 0$~K,
2 -- $E = 1 \cdot 10^{-2}$~K, 3 -- $E = 2 \cdot 10^{-2}$~K, 4 -- $E = 4 \cdot 10^{-2}$~K.
}
%%\label{prsfig2}
\end{figure}

The field dependence of $\chi$ for $\tau>0$ is well defined, all curves are symmetric with respect to $E=0$. The field dispersion of $\chi$ is proportional to $\tau$
[figure~5~(a)].
\begin{figure}[!ht]
\includegraphics[width=0.48\textwidth]{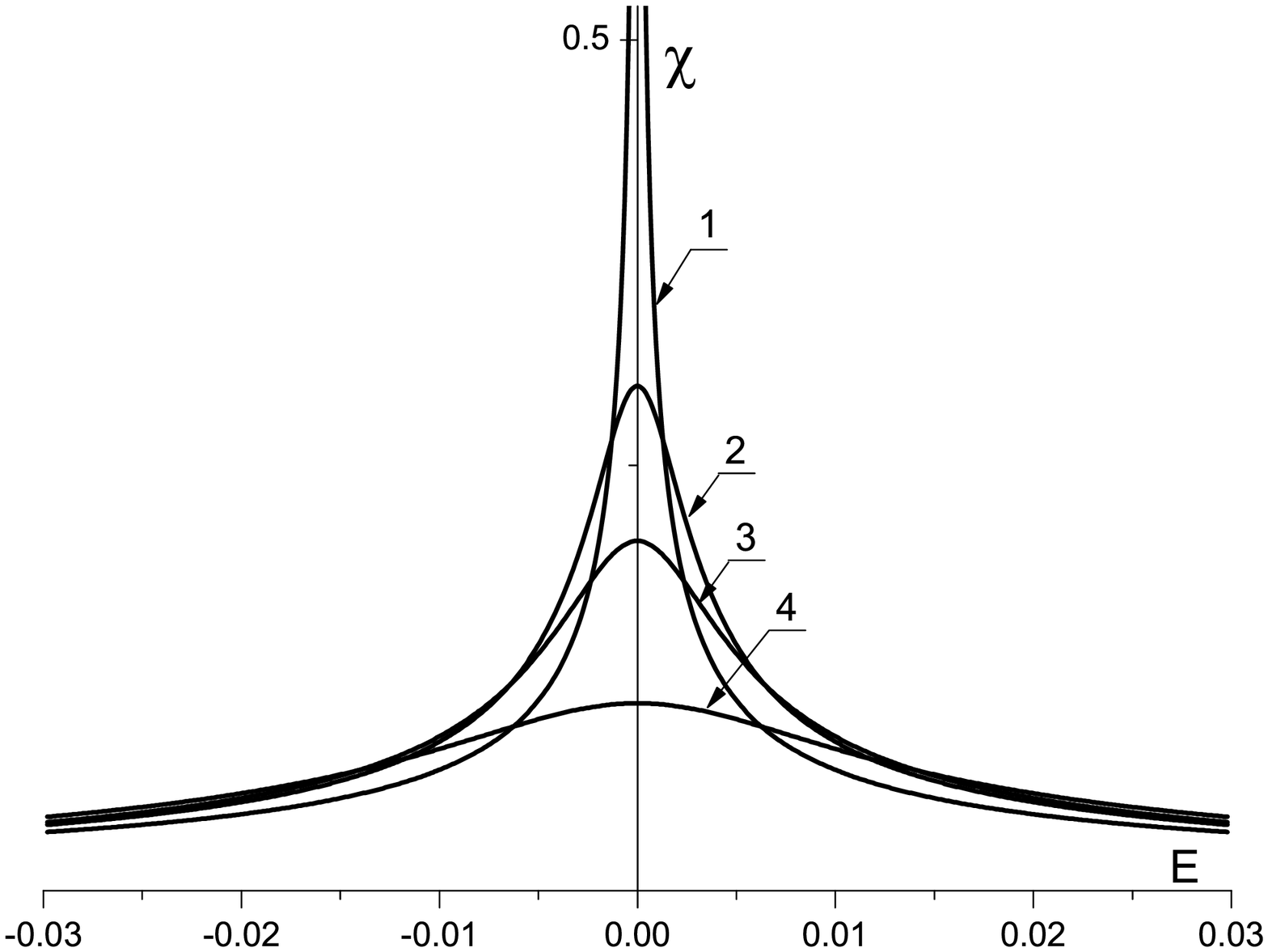}%
\hfill%
\includegraphics[width=0.48\textwidth]{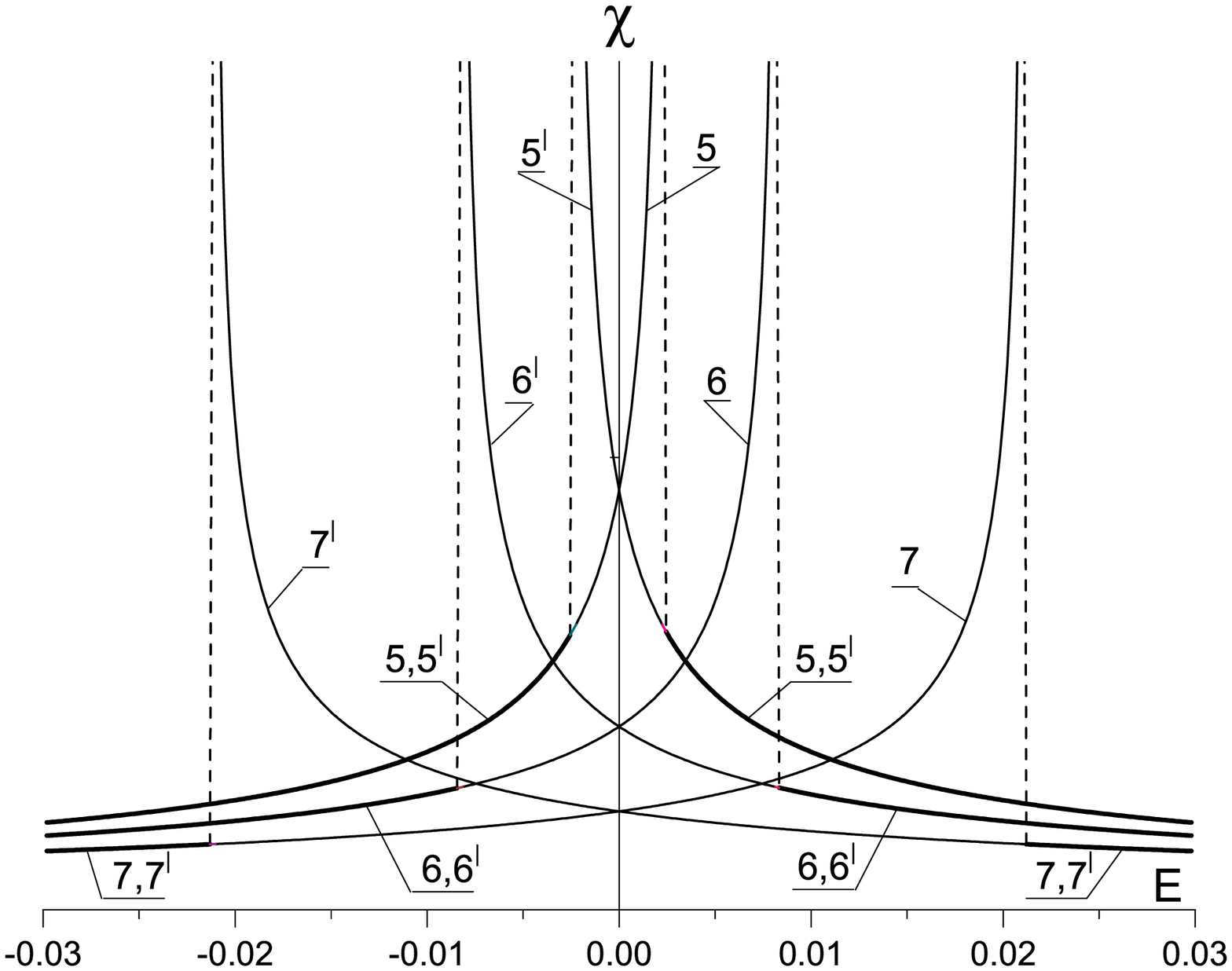}%
\\%
\parbox[t]{0.48\textwidth}{%
\centerline{(a)}%
}%
\hfill%
\parbox[t]{0.48\textwidth}{%
\centerline{(b)}%
}%
\caption{Field dependencies of the static dielectric ``susceptibility'' for $\tau>0$ (a) and $\tau<0$ (b):
1 -- $\tau = 1\cdot 10^{-6}$,
2 -- $\tau = 2 \cdot 10^{-5}$, 3 -- $\tau = 3 \cdot 10^{-5}$, 4 -- $\tau = 6 \cdot 10^{-5}$,
5 -- $\tau = -2 \cdot 10^{-5}$, 6 -- $\tau = -5 \cdot 10^{-5}$, 7 -- $\tau = -1 \cdot 10^{-4}$.
Curve numbers $5'$, $6'$, $7'$ correspond to the oppositive change (from positive values to negative ones) of the external field $h$.
Solid lines (b) describe the behaviour of $\chi$ independent of the direction of $h$ change.}
\end{figure}

However, for $\tau<0$, the situation is much more complicated [figure~5~(b)]. The position of $\chi$ maxima (at different $\tau$) depends on the direction of the external field $E$
change. When $E$ arises (from negative values to positive ones) the $\chi$ maximum is located in the area of positive $E$.  In the opposite case, ($E$ changes from
positive values to negative ones) the $\chi$ maximum is located in the area of negative $E$. This behavior is in strict correlation with the hysteresis of $\rho$
(see figure~3).

The temperature behaviour of the order parameter $\rho$ and static dielectric susceptibility $\chi$ with logarithmic corrections makes it possible to obtain the effective temperature critical indices $\beta^*$ and $\gamma^*$ from the relations:
\be
\rho \sim |\tau|^{\beta^*}, \qquad
\chi \sim |\tau|^{-\gamma^*}.
\ee

Correspondingly, effective field critical indices $\delta^*$ and $\epsilon^*$
\be
\rho \sim |E|^{1/\delta^*}, \qquad \chi \sim |E|^{-\epsilon^*}
\ee
can be also calculated.

For example, at $\tau=10^{-5}$, we obtain:
\[
\beta^* = 0.4552, \qquad \gamma^* = 0.9121, \qquad \delta^* = 3.27, \qquad {\cal{E}}^* = 0.69.
\]
Inasmuch as for $E=0$ the equation (4.2) can be solved in the analytic form:
\be
\rho_1 = -\rho_2 = \sqrt{\frac{B}{2D}}, \qquad \rho_3 = 0,
\ee
the effective critical indices $\beta^*$ and $\gamma^*$ with logarithmic corrections can be easy calculated. From the comparison of (4.4), (4.6)
and their derivatives we have:
\be
\beta^* = \frac{1}{2} + \frac{1}{6\ln|\tau|}\,,\qquad
\gamma^* = 1 + \frac{1}{3\ln|\tau|}\,,
\ee
so in the limit $\tau\rightarrow 0$, their classic values take place. However, the logarithmic convergence is very weak.

It must be noted that due to dipole--dipole (anisotropic) and long-range interaction, the critical behaviour of the system investigated is wholly different from the
isotropic Ising model critical behaviour \cite{13,14}, where strong non-classical exponents take place.
Due to the specific recursion relations (3.15) the fixed point for $r_n$, $u_n$ (3.16) is of Gaussian type and the conclusion on
the classical behaviour is obvious. Outside the immediate neighbourhood of $T_{\mathrm{c}}$, the system behaves similar to the isotropic
one (the fine characteristics of interparticle potential do not manifest themselves  decisively) and effective critical indices $\beta^*$
and $\gamma^*$ are somewhat less than its limit ($\tau\rightarrow 0$) values. Due to this fact, the effective critical index $\alpha^* = -2/(3\ln|\tau|)>0$
and specific heat behaviour near $T_{\mathrm{c}}$ is in agreement with the experimental data~\cite{15}.

The main peculiarity of the obtained results is a very weak dependence of $\beta^*$ and $\gamma^*$ on the dividing parameter $S$, while in the isotropic
Ising model, the corresponding dependence is great. The latter is one of unsolved problems in the theory using CVM \cite{1,2}.
When $|T\rightarrow T_{\mathrm{c}}|$, the susceptibility increases
weakly as compared with its behaviour in the self-consistent field approximation $(\gamma = 1)$. The ``law of doubling''
($\chi$ for $T>T_{\mathrm{c}}$ is twice bigger than $\chi$ for $T<T_{\mathrm{c}}$) has been fulfilled.

Thus, the basic properties of the investigated ferroelectric cluster system are described.
The order parameter is calculated and its critical behaviour is analyzed.

\section{Conclusions}

An essential generalization of CVM towards investigation of Ising-like systems with non-isotropic interactions near the second order phase transition point has been
proposed. The basic distribution for CV contains an external field $E$ and the higher order terms as compare to Gaussian term  and depends on a dipole--dipole
interparticle interaction. As a result, the obtained physical characteristics of the studied ferroelectric cluster system demonstrate non-classical
behaviour near $T_{\mathrm{c}}$ with effective critical indices.

The main advantage of the CVM used herein, as compared with the usual renormalization group methods, is the possibility to calculate an order parameter and to describe its behaviour near the critical point ($T=T_{\mathrm{c}}$ and $E=0$). A complete set of both temperature and field critical dependencies are
obtained here. The logarithmic corrections to critical indices are calculated for the first time in counterbalance to \cite{12}, where similar corrections were applied to
temperature dependencies of the order parameter and dielectric susceptibility only.
The power of temperature logarithmic correction to dielectric susceptibility is twice bigger than the power to the order parameter while
in \cite{12} they are declared as equal each other.

A scrupulous study of the effect of the external field on the critical behaviour of the order parameter and on dielectric susceptibility of the ferroelectric system confirms its
suppressive action, the shift of the point of transition and hysteresis of the order parameter behaviour. The position of the points of the first order phase transition depends on the direction of the field change.

\ukrainianpart

\title{Сегнетоелектричний параметр порядку двочастинкової кластерної системи поблизу точки фазового переходу. Метод колективних змінних}

\author[М.А. Кориневський, В.Б. Солов'ян]{М.А. Кориневський\refaddr{label1,label2,label3},
        В.Б. Солов'ян\refaddr{label1}}
\addresses{
\addr{label1} Інститут фізики конденсованих систем НАН України, вул. Свєнціцького 1, 79011 Львів, Україна
\addr{label2} Національний університет ``Львівська політехніка'' вул. С.~Бандери 12, 79013 Львів, Україна
\addr{label3}  Щецінський університет, Інститут фізики, вул. Вєлькопольська 15, 70451 Щецін, Польща
}

\makeukrtitle

\begin{abstract}
\tolerance=3000%
Запропоновано новий підхід для вивчення поведінки параметра впорядкування поблизу точки сегнетоелектричного фазового переходу. Враховано короткосяжні і дипольні взаємодії
між частинками. Розраховано і обговорено логарифмічні поправки та ефективні критичні показники.
\keywords фазовий перехід, параметр порядку, критичні показники, метод колективних змінних
\end{abstract}

\end{document}